\documentstyle{article}
\begin{document}
\title{On Continuous Ambiguities in Model-Independent Partial Wave
Analysis - II.}
\author{I.N.Nikitin* }
\date{}
\maketitle
\insert\footins{\ \ {\footnotesize *  E-mail:\ \ nikitin\_{$\,$}i@mx.ihep.su }}
\def\n{\vec n}
\def\O{\Omega}
\def\dO{\partial\Omega}
\def\ph{\varphi}
\def\half{{\textstyle{{1}\over{2}}}}
\def\rat#1#2{{\textstyle{{#1}\over{#2}}}}
\def\rea{\mbox{Re}\: }
\def\img{\mbox{Im}\: }
\def\intn{\int d^{2}n\ }
\def\la{\Bigl}
\def\ra{\Bigr}
\def\ket#1{|#1\ra>}
\def\bra#1{\la<#1|}
\def\twin#1#2{\scriptstyle #1\atop\scriptstyle #2}
\def\no{\nonumber}

\begin{abstract}
A problem of density matrix determination in terms of the
given angular distribution of decay products is considered.
\end{abstract}

In the first part of this paper (Section III) a quantum system, described
by density matrix, has been considered. In this the density matrix was diagonal
with respect to orbital quantum numbers and the state was mixed only partially.
Now we will consider the process $X\to0+0$, where initial state is described by
the density matrix $\rho$, for which we do not assume any other properties
except hermitity and positive definiteness.

The angular distribution is given by the expression
\begin{equation}
I(\n)=\sum_{\alpha\beta}\rho_{\alpha\beta}Y_{\alpha}^{*}(\n)Y_{\beta}(\n).
\label{Irho}
\end{equation}
Let us multiply both sides of this expression by $Y_{\gamma}(\n)$ and
integrate this over the sphere
\begin{eqnarray}
&&t_{\gamma}=\sum_{\alpha\beta}\rho_{\alpha\beta}D_{\alpha\beta\gamma},
\label{trD}\\
\mbox{where}\quad&&t_{\gamma}=\intn IY_{\gamma}=<Y_{\gamma}>_{I}
\mbox{\ are moments of distribution,}\no\\
&&D_{\alpha\beta\gamma}=\intn Y_{\alpha}^{*}Y_{\beta}Y_{\gamma}.\no
\end{eqnarray}
The integrals $D_{\alpha\beta\gamma}$ are expressed in terms of Clebsch-Gordan
coefficients (see \cite{Landau}\S 107):
$$D_{l_{1}m_{1},l_{2}m_{2},l_{3}m_{3}}=(-1)^{m_{3}}
\sqrt{{{(2l_{1}+1)(2l_{3}+1)}\over{4\pi(2l_{2}+1)}}}
\ C_{l_{1}\ -m_{1},l_{3}m_{3}}^{l_{2}\ -m_{2}}\ C_{l_{1}0,l_{3}0}^{l_{2}0}\ .
$$
We use spherical functions \cite{PDG}, different from those of \cite{Landau} by
a phase factor:  $Y_{lm}^{\mbox{{\small\cite{Landau}}}}=
i^{l}Y_{lm}^{\mbox{{\small\cite{PDG}}}}$.
Summation in (\ref{trD}) extends for the following values of indices:
$$m_{2}=m_{1}-m_{3},\quad l_{2}=|l_{1}-l_{3}|\ ...\ l_{1}+l_{3}\
\ \mbox{step 2},$$
$$-\min\{ l_{1},l_{2}-m_{3}\}\leq m_{1}\leq\min\{ l_{1},l_{2}+m_{3}\} .$$
Let us represent the density matrix $\rho_{\alpha\beta}$ via eigenvalues
$\lambda_{i}$ and eigenvectors~$\Psi_{\alpha}^{i}$
$$\rho_{\alpha\beta}=\sum_{i}\lambda_{i}\Psi_{\alpha}^{i*}\Psi_{\beta}^{i}.$$
The $\Psi_{\alpha}^{i}$ form unitary matrix
$$\sum_{i}\Psi_{\alpha}^{i*}\Psi_{\beta}^{i}=\delta_{\alpha\beta}\ .$$
The positivity of $\rho$ implies that all eigenvalues are positive:\
$\lambda_{i}>0$.\\
Let us rewrite (\ref{trD})
\begin{eqnarray}
t_{\gamma}&=&\sum_{i}M_{\gamma i}\lambda_{i},\label{tMl}\\
M_{\gamma i}&=&\sum_{\alpha\beta}\Psi_{\alpha}^{i*}\Psi_{\beta}^{i}
D_{\alpha\beta\gamma}.\no
\end{eqnarray}
At $\gamma=0$ from (\ref{tMl}) a normalization condition follows:
$$\sum_{i}\lambda_{i}=\intn I=1,\ \mbox{ the distribution is normalized to
unity.}$$
At the given $\Psi_{\alpha}^{i}$ relation (\ref{tMl}) is a system of linear
equations for $\lambda_{i}$. A matrix of this system is non-degenerate,
except rare special values of $\Psi_{\alpha}^{i}$.
\begin{quotation}{\small
Particularly, the matrix $M_{\gamma i}$ degenerates when $\Psi_{\alpha}^{i}=
\delta_{i\alpha}$. The question whether the matrix $M_{\gamma i}$ is
non-degenerate reduces to clarification of linear independence of a set of
functions $|\Psi_{i}(\n)|^{2}$ for the given complete set of functions
$\Psi_{i}(\n)$.
}\end{quotation}

Therefore in our approach the free parameters are the unitary matrices
$\Psi_{\alpha}^{i}$, the values $\lambda_{i}$ are expressed via
$\Psi_{\alpha}^{i}$ from system (\ref{tMl}), the inequalities
$\lambda_{i}(\Psi)>0$ select a region of possible values for $\Psi$.

\underline{Notion.}

The presence of continuous ambiguity in the
solution of this problem is most obvious
in $\n$-representation of operators. The density matrix is represented by the
kernel of integral operator
$$(\rho\Psi)(\n)=\int d^{2}n'\rho(\n,\n ')\Psi(\n ').$$
The distribution is equal to the diagonal elements
\begin{equation}
I(\n)=\rho(\n,\n)\label{Inn}
\end{equation}
Hence, the problem is reduced to the reconstruction of hermitean positive
matrix
from the given diagonal. Let $\rho$ be a solution of this problem. The
replacement $\rho\to\rho+\Delta\rho$, where $\Delta\rho$ is arbitrary hermitean
matrix with zero diagonal elements, conserves (\ref{Inn}) and hermitity of
$\rho$. If the matrix elements of $\Delta\rho$ are sufficiently small, then the
positivity of $\rho$ is also conserved. Therefore, the
solution of the problem is not unique.

In $\n$-representation the analysis of $\rho$ positivity and the determination
of eigenvectors are complicated. The approach proposed above seems more
convenient.

\subsection*{Scheme of analysis}

When describing quantum systems via density matrix one suggests that
the system is located in pure states $\Psi_{\alpha}^{i}$ with different
probabilities $\lambda_{i}$. The representation of eigenvectors of density
matrix on Argand plots is an ultimate goal of partial wave analysis. The input
data are the moments $t_{\gamma}$. We propose to obtain the region of solution
using Monte-Carlo technique. The scheme of analysis is the following:
\begin{enumerate}
\item
A random unitary matrix $\Psi_{\alpha}^{i}$ is generated. We use the generation
of the columns of this matrix as a set of independent isotropically distributed
random complex vectors with the subsequent Hilbert-Schmidt orthogonalization.

An isotropic (Gaussian) distribution of $n$-dimensional vector can be obtained
by generating its components as independent random values with Gaussian
distribution:
$$dp=e^{-{x_{1}}^{2}}dx_{1}\ ...\ e^{-{x_{n}}^{2}}dx_{n}=e^{-r^{2}}dV.$$
The unitary matrices generated by this me\-thod are uni\-form\-ly
di\-stri\-bu\-ted over
the unitary group invariant volume\footnote{The unitary transformation
transfers
one orthonormal basis to another. The invariant measure on unitary group can be
defined as a product of spherical volume for one vector from the
basis and the measure on subgroup leaving this vector invariant (little
subgroup). Using this recurrent definition and isotropy of generated vectors,
one can easily prove the above statement by induction.}.
\item
The matrix $M_{\gamma i}$ is obtained, system (\ref{tMl}) is solved for
$\lambda_{i}$.
\item
The solution is accepted, if all $\lambda_{i}>0$, and is discarded otherwise.
\item
The eigenvalues $\lambda_{i}$ are displayed on intensity plots, the
eigenvectors
$\Psi_{\alpha}^{i}$ -- on Argand plots.
\end{enumerate}

\underline{Remarks.}

1) Phase rotation of eigenvectors $\Psi_{\alpha}^{i}\to e^{i\ph_{i}}
\Psi_{\alpha}^{i}$ does not change the density matrix. One can generally choose
the phases of eigenvectors in such a way that the component $\Psi_{0}^{i}$
in each eigenvector will be a real positive number (phase is measured from
S-wave). Those unitary $N\times N$ matrices are described by $N^{2}-N$
independent real parameters, which being unified with $N$ eigenvalues
$\lambda_{i}$ give $N^{2}$ degrees of freedom, contained in an arbitrary
hermitean matrix $\rho_{\alpha\beta}$.

2) Change of numeration (permutation) of eigenvalues and eigenvectors
conserves the density matrix. The region of solutions is symmetrical with
respect to these permutations. Thus projections of the region onto Argand
plots of each eigenvector coincide. One can order the eigenvalues:
$\lambda_{1}\geq\lambda_{2}\geq...\geq\lambda_{N}$ and respectively permute
the eigenvectors. Such ordering might lead to breaks of continuity in
solutions.
The breaks appear when a collision of eigenvalues occurs on a smooth solution
(fig.1). The ordering transfers the continuous path $ab$ to the non-continuous
path $ab'$. One should take into account this fact when selecting the
solutions: not only smooth paths inside ambiguity region are allowed,
but also those paths, which can be smoothly continued in the points with
coincident eigenvalues after the permutation of the correspondent eigenvectors.
\begin{quotation}{\small
In the vicinity of the point $A$ the eigenvectors $\Psi_{1}$ and $\Psi_{2}$
cannot be close because they are orthogonal. The points $A$ and $A'$ cannot
coincide on all Argand plots. Hence the break of curves on Argand plots
actually
occurs in transition through the point $A$ along the path $ab'$. One can
connect
the points $A$ and $A'$ by continuous path, along which the density matrix is
constant. At $\lambda_{1}=\lambda_{2}$ any linear combinations of $\Psi_{1}$
and
$\Psi_{2}$ are also eigenvectors. The $\Psi_{1}$ and $\Psi_{2}$ can be permuted
with the aid of unitary transformation, not touching others $\Psi_{i}$ (this
can be developed continuously, because $U(2)$ group is connective). The
rotation
$AA'$ is developed at constant $\lambda_{1}=\lambda_{2}$. Hence, though the
transition $aAA'b'$ is continuous, it cannot have analytical dependence on
energy.)
}\end{quotation}

3) Let first $N$ moments of distribution be measured in experiment
($l\leq L_{max},\ N=(L_{max}+1)^{2}$). Let the unitary matrices $\Psi$
have sizes $N\times N$. We will solve $N$ first equations of (\ref{tMl})
for $N$ unknowns $\lambda_{i}$. The density matrices obtained in solution
give the distributions, for which $N$ first moments exactly coincide with
measured ones. Other moments $t_{\gamma},\ \gamma >N$ can be obtained by
substitution of values $\Psi_{\alpha}^{i} ,\lambda_{i}$ into (\ref{tMl}).
Generally these moments do not vanish. One can show from the properties of the
coefficients $D_{\alpha\beta\gamma}$ that there is a finite number of non-zero
moments: $L_{max}<l\leq2L_{max}$.

The partial wave analysis
practice is to treat higher (non-measured) moments as zeros
with the experimental precision. One can impose conditions on these moments
$$|t_{lm}|<\delta t,\qquad  L_{max}<l\leq2L_{max},$$
where $\delta t$ is a statistical error of the moment.
$$|\delta t_{\gamma}|^{2}={{DY_{\gamma}}\over{N_{\mbox{\small evts}}}},$$
\quad$N_{\mbox{\small evts}}$
is the total number of events, on which the moment is measured
$$t_{\gamma}={{1}\over{N_{\mbox{\small evts}}}}
\sum\limits_{i=1}^{N_{\mbox{\small evts}}}
Y_{\gamma}(\n_{i}),$$

$DY_{\gamma}$ is the dispersion of random value $Y_{\gamma}(\n)$,
$$DY_{\gamma}=<|Y_{\gamma}|^{2}>-\ |<Y_{\gamma}>|^{2}=\intn I|Y_{\gamma}|^{2}\
-
|t_{\gamma}|^{2},$$
$$|\delta t_{\gamma}|^{2}\leq {{1}\over{N_{\mbox{\small evts}}}}
\intn I|Y_{\gamma}|^{2}.$$
Averaging both sides over the quantum number $m$, we have
$${{1}\over{2l+1}}\sum_{m}|\delta t_{lm}|^{2}
\leq{{1}\over{4\pi N_{\mbox{\small evts}}}},\quad
\mbox{i.e.\ } \overline{|\delta t_{lm}|}\leq{{1}
\over{\sqrt{4\pi N_{\mbox{\small evts}}}}}
=\delta t.$$
When $|t_{\gamma}|>\delta t$, the deviation of $t_{\gamma}$ from zero is
statistically considerable.

We note, that our estimate for $\delta t$ is $l$ independent, all moments are
measured with equal precision. Actually one can reliably measure a great number
of moments (limitations are imposed only by finite angular distribution of the
equipment: $l<2\pi/\Delta\theta_{\mbox{\small resol}}$).

The contributions of the moments with high $l$ in distribution $I(\n)$ are
fast oscillating functions. To detect $l$-th harmonics one should fill 2D
histograms, containing at least $l^{2}$ bins. Due to limited statistics
a few events will be placed in one bin. Hence, the presence of high harmonics
can not be detected in angular distribution. Nevertheless, the moments of these
harmonics can be precisely measured.

\vspace{1cm}

\underline{Example.}\quad Let us consider the distribution
\begin{equation}
I(\n)=(1-a)|Y_{0}|^{2}+a|Y_{10}|^{2}={{1}\over{4\pi}}
(1-a+3a\cos^{2}\theta).\label{Ia}
\end{equation}
The moments are
\begin{equation}
t_{0}={{1}\over{\sqrt{4\pi}}},\quad
t_{20}={{a}\over{\sqrt{5\pi}}},\quad
\mbox{others\ }t_{\gamma}=0.\label{ta}
\end{equation}
Let first 9 moments $t_{\gamma}$ be measured in an experiment
(S, P and D-moments)\footnote{The distribution can differ from (\ref{Ia})
by the contributions of higher moments $l\geq3$.}. The density matrix is
described by 81 real parameters, 72 of which describe the unitary matrix
$\Psi$ ($\Psi_{0}^{i}\in R_{+}$), 9 eigenvalues $\lambda_{i}$ can be obtained
from system (\ref{tMl}).

Figures 2 and 3 show the result of analysis for the value $a=1/6$. The number
of random unitary matrices was 100000. In 1646 cases the solution of system
(\ref{tMl}) satisfied the condition $\lambda_{i}>0$. For 54 points the
additional condition $|t_{lm}|<10^{-2},\ l\geq3$ held true.

The solutions $\Psi$ fill a region in the group $U(9)$. In projection onto
Argand plots this region maps into a set of related points.  The density of
points on figures estimates a ``thickness'' of the stratum, projected
into the same area on the plot. A circle on the figures denotes the trivial
solution
\begin{eqnarray}
\lambda_{1}&=&5/6\quad\Psi_{0}^{1}=1,\quad\mbox{others\ }\Psi_{\alpha}^{1}=0
\quad\mbox{(S-wave)}\no\\
\lambda_{2}&=&1/6\quad\Psi_{10}^{2}=1,\quad\mbox{others\ }\Psi_{\alpha}^{2}=0
\quad\mbox{(P-wave)}\no\\
\lambda_{i}&=&0,\ i=3..9\no
\end{eqnarray}
A few points fall in the vicinity of the trivial solution, most part
of solutions is located in the region of close eigenvalues
$\lambda_{i}=0...0.2$.

The points, for which the additional condition $|t_{lm}|<10^{-2},\ l\geq3$
is satisfied, cover the same areas on the figures. A position of these
solutions
is analogous to the position of thin layer in a sphere: outer spherical layer
occupies a small volume in a sphere, but covers the same area in projection
on a plane.

\vspace{0.5cm}

\underline{Limiting cases.}

\underline{$a\to0$}\quad
Isotropic distribution ($a\to0$) is a singular case for the given scheme of
analysis. When $t_{\gamma}=\delta_{\gamma0}/\sqrt{4\pi}$, a set of equal
eigenvalues $\lambda_{i}=1/N$ is a solution of (\ref{tMl}) for all $\Psi$.
In this the density matrix is proportional to unit matrix: $\rho_{\alpha
\beta}=\delta_{\alpha\beta}/N$, the isotropy of distribution (\ref{Irho})
follows from the property $\sum\limits_{m=-l}^{l}|Y_{lm}(\n)|^{2}=(2l+1)/4\pi$.
The solution $\lambda_{i}=1/N$ has no definite limit at $N\to\infty$ and it
is not of physical value. Also there are other solutions, e.g. pure S-wave
state
or $$ \Psi=\left(\begin{array}{cccc}
1&0&0&\ldots\\
0&U(3)&0&\ldots\\
0&0&U(5)&\ldots\\
\ldots&\ldots&\ldots&\ldots\\
\end{array}\right)\quad ,\quad
\Lambda=\left(\begin{array}{cccc}
\lambda_{1}&0&0&\ldots\\
0&\lambda_{2}\cdot1_{3\times3}&0&\ldots\\
0&0&\lambda_{3}\cdot1_{5\times5}&\ldots\\
\ldots&\ldots&\ldots&\ldots
\end{array}\right)\quad ,
$$
the matrix $\Psi$ is diagonal with respect to $l$ (mixes $m$ only),
correspondent $\lambda_{lm}$ are independent of $m$. System (\ref{tMl})
has these additional solutions only if the matrix $M_{\gamma i}$ is degenerate.
The values of parameters $\Psi$, for which $M_{\gamma i}$ degenerates, form
in unitary group a set of zero measure. When generating
matrices $\Psi$ uniformly
distributed in $U(N)$ volume these additional solutions will not be revealed.

Transition to the limit $a\to0$ is shown on fig.4. The distribution of
$\lambda_{i}$ contracts to the point $\lambda=1/9\ (N=9)$, the points on Argand
plots fill unit circles.

The presence of such singularities indicates that the whole set of solutions
of system (\ref{tMl}) possesses a complicated topological structure.
This structure can be inferred from the following examples (fig.5):

$$\mbox{a)}\quad
\left(\begin{array}{cc}1&1\\x&y\end{array}\right)
\left(\begin{array}{c} \lambda_{1}\\ \lambda_{2}\end{array}\right)=
\left(\begin{array}{c} 1\\ 0\end{array}\right)\quad\Rightarrow\quad
\left(\begin{array}{c} \lambda_{1}\\ \lambda_{2}\end{array}\right)=
{{1}\over{y-x}}\left(\begin{array}{c} y\\ -x\end{array}\right).
$$
At $x=y=0$ the solutions of the system (with positive $\lambda_{i}$) form
a segment
$(0<\lambda_{1}<1,\ \lambda_{2}=1-\lambda_{1})$. This set has greater dimension
($d=1$, a line) than the set of solutions at fixed $x\neq y$ ($d=0$, a point),
but less dimension than the set of solutions for all $x,y$ ($d=2$,
a surface). Singular solutions have zero measure in the set of all solutions
(on the hyperbolic paraboloid, fig.5a).
$$\mbox{b)}\quad
\left(\begin{array}{ccc}1&1&1\\x&-x&0\\y&0&-y\end{array}\right)
\left(\begin{array}{c} \lambda_{1}\\ \lambda_{2}\\ \lambda_{3}\end{array}
\right)=
\left(\begin{array}{c} 1\\ 0\\ 0\end{array}\right).
$$
At fixed $x\neq0,y\neq0$ the solution is a point $\lambda_{i}=1/3$.
When $x=0$ or $y=0$, the dimension of the set of solutions is increased by 1,
when $x=y=0$ -- by~2. The set of singular solutions has the same dimension as
the set of all non-singular ones ($d=2$).

In uniform random generation of parameters $(x,y)$ the singular solutions will
not be found. This is essential only for singularities of the type~b).

\vspace{0.5cm}

\underline{Limits of $a$.}\quad
Function (\ref{Ia}) is positive on the sphere at $-1/2<a<1$. For these
values of $a$ the problem considered has solutions.

Remarks.

1) At negative $a$ the point $\{ \lambda_{1}=1-a,\ \Psi_{1}=\mbox{S-wave};\
\lambda_{2}=a,\ \Psi_{2}=\mbox{P-wave}\}$ is no longer a positively defined
solution of the problem.

2) Non-negativity of the distribution is necessary and sufficient condition
for positively semi-defined density matrix existence, if sizes of matrices
are not bonded ($N\to\infty$). The necessity is obvious, the sufficience
follows from the fact that any distribution $I(\n)\geq0$ can be described by
a pure state (see first part of this paper, Section I), i.e. the density matrix
with single non-zero eigenvalue
\begin{equation}
\lambda_{1}=1,\ \lambda_{i}=0, i>1,\ \Psi_{1}(\n)=\sqrt{I(\n)}e^{i\ph(\n)}.
\label{pur}
\end{equation}

3) At finite $N$ this condition is neither necessary nor sufficient. We study
a class of distributions which can differ by harmonics with numbers greater
than
$N$. Even if $I(\n)<0$ in some point for distribution considered, positive
functions can exist in the same class. On the other hand, the pure states of
form (\ref{pur}) have in general an infinite number of harmonics and are not
contained in the considered finite dimensional class $\Psi_{\alpha}^{i}$.
One can show that a set of $a$ values, for which the problem has solutions in
the finite dimensional class, is connective, i.e. it is a segment
$a_{-}\leq a\leq a_{+}$, with $a_{+}\geq1$ for distribution (\ref{Ia}).
We will not determine exact values $a_{\pm}$.

When $a$ tends to boundary points, the volume occupied by solutions in $U(N)$,
tends to zero fast. The behaviour of this volume could be inferred from the
data
given in Table~1.

\begin{table}
\caption{ Number of positively defined solutions for 50000 random matrices with
random distribution in $U(9)$ }
 \begin{center}\begin{tabular}{|c||c|c|c|c|c|c|c|c|}\hline
$a$&-1/4&-1/8&-1/16&0&1/16&1/8&1/6&1/4\\ \hline
$N_{\mbox{\small sol}}$&9&2192&14689&50000&14969&2770&823&75\\
\hline
 \end{tabular}\end{center}
\end{table}

At large $|a|$ and at a greater number of moments involved in analysis as well,
the Monte-Carlo based approach presented here is ineffective
(most part of solutions is discarded in positivity test $\lambda_{i}$).
One might use more advanced methods for solution of inequalities system
$\lambda_{i}(\Psi)>0$.

\subsection*{Discussion}
The pure states are particular cases of mixed states, hence the ambiguity
region
for mixed states is wider than for pure ones. Even in the class of
solutions, containing a
finite number of harmonics, continuous transformations of
density matrix are available, which do not change the distribution.
One can fix the continuous ambiguity of solutions only imposing model
restrictions on the form of density matrix. The illustration of this statement
can be found in \cite{Hansen}. This work presents a formalism used in partial
wave analysis of the reactions $\pi p\to(3\pi)p,\ Kp\to(K\pi\pi)p$. The process
is represented in a form of decay sequence\footnote{It is assumed that the
process amplitude has a general form and depends on all quantum numbers and
kinematic variables. In this step the representation of the process as decay
sequence is introduced for convenience and do not impose any restrictions on
the
form of amplitude.}
$$\mbox{ meson }+ p\to p+\mbox{ meson }(J^{P});$$
$$\mbox{ meson }(J^{P})\to\mbox{ meson }+\mbox{ di-meson };$$
$$\mbox{ di-meson }\to\mbox{ 2 mesons }.$$
Then assumptions for the amplitude (eigenvectors of density matrix) are made,
of which the most important are:

1. The amplitude dependence on kinematic variables, describing the decay of
di-meson, is introduced as a product of Breit-Wigner function and barrier
factors.

2. The amplitude of the process is represented as a product of $J^{P}$ state
production amplitude and its decay amplitude. This assumption is exact only if
a single state $J^{P}$ is present.

As stated in the work, namely these assumptions fix continuous ambiguity of
partial wave analysis (1--partially, 2--completely).

\vspace{0.5cm}

The author is indebted to E.B.Berdnikov, I.A.Kachaev, D.I.Ryabchikov and
S.A.Sadovsky for helpful discussions.

\vspace{0.5cm}

\hfill {\it Received December 8, 1994.}
\newpage
\subsection*{Figure captions}
\parindent=0cm

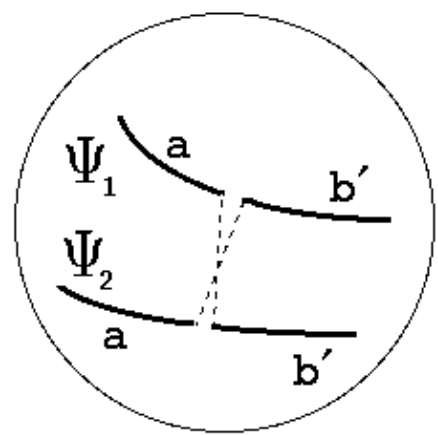
{\bf Fig.1.\ }
In point $A$ the eigenvalues $\lambda_{1}$ and $\lambda_{2}$ coincide.
The continuous path $ab$ transfers after ordering of eigenvalues into
the path $ab'$, lying in the region $\lambda_{1}>\lambda_{2}$ (hatched on
fig.b). The eigenvectors $\Psi_{1}$ and $\Psi_{2}$ can not coincide
(the case (c) is impossible). In the point $A$ the break of curves on Argand
plot occurs (fig.d).
\vspace{0.5cm}

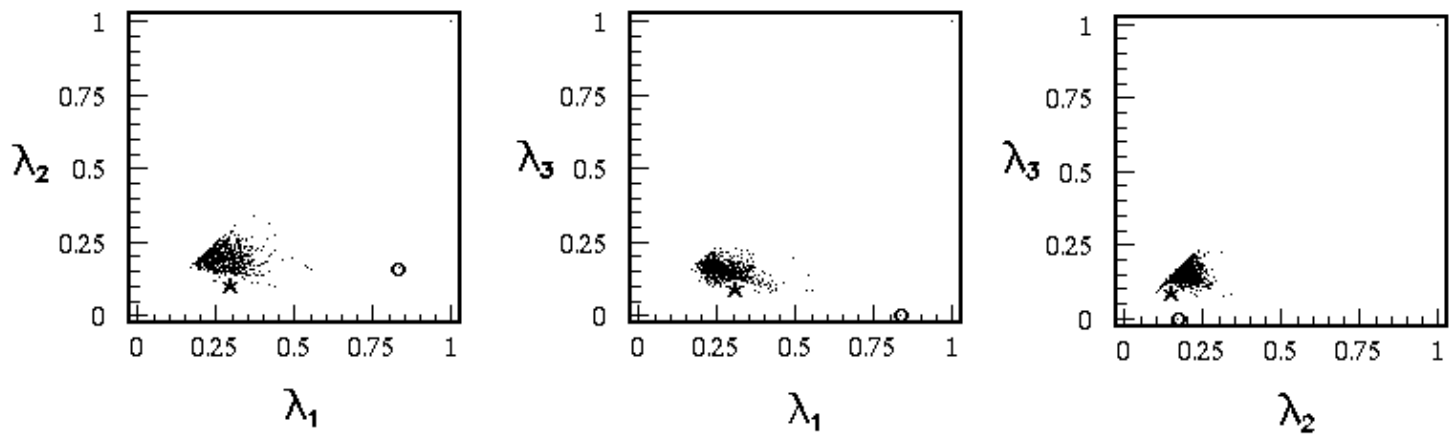
{\bf Fig.2.\ }
Ambiguity region: eigenvalues. A circle denotes the trivial solution (mixture
of S and P-waves). One arbitrarily selected solution is denoted by a star.
\vspace{0.5cm}

{\bf Fig.3.\ }
Ambiguity region: eigenvectors. a) $\Psi_{1}$;\ b) $\Psi_{2}$.
\vspace{0.5cm}

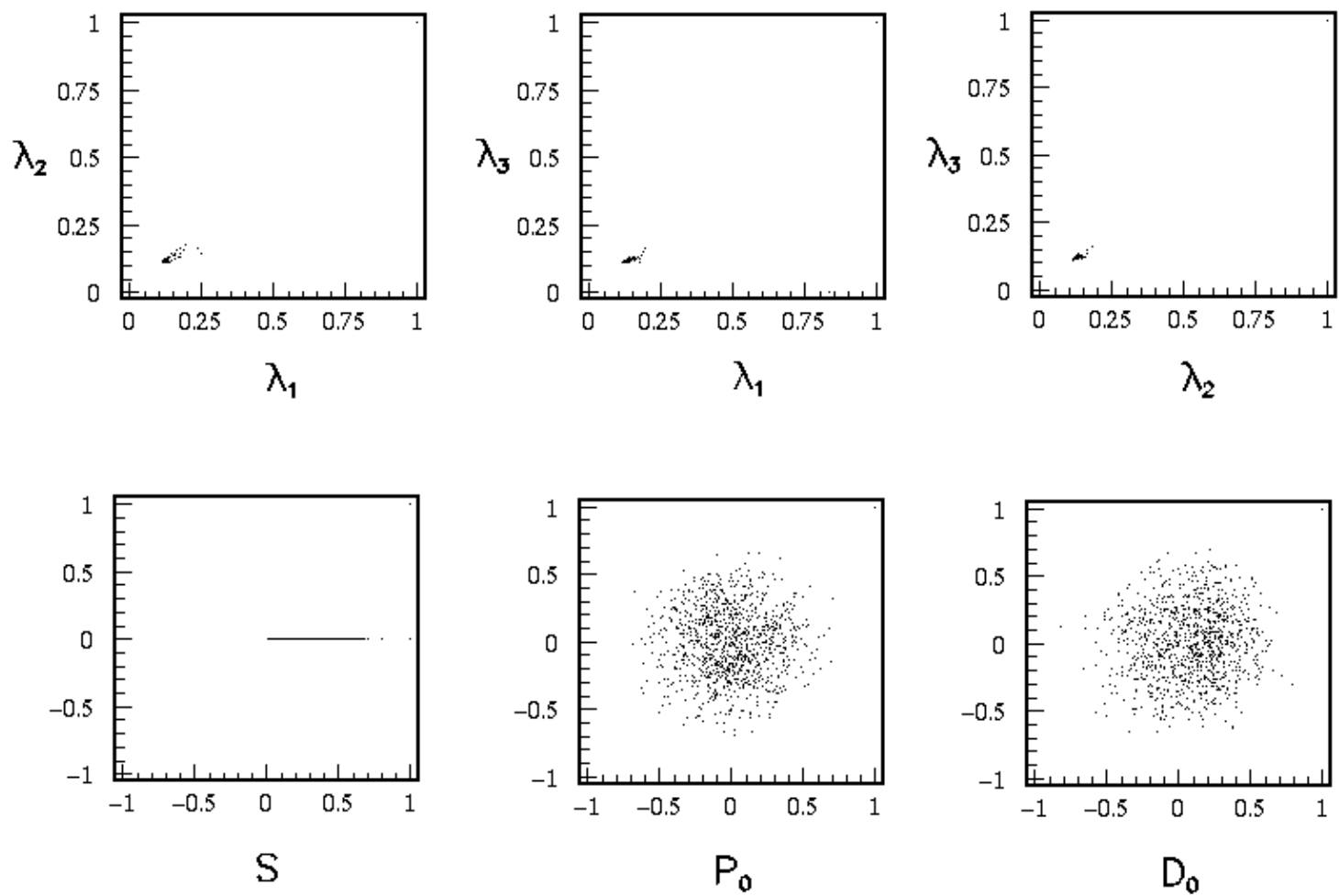
{\bf Fig.4.\ }
Ambiguity region: $a\to0$.
\vspace{0.5cm}

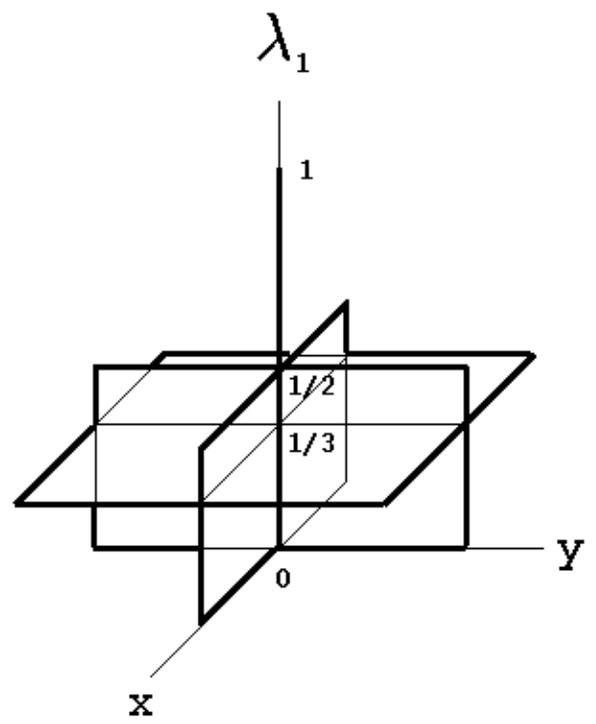
{\bf Fig.5.\ }
Parametric dependence of linear systems solutions (see text).

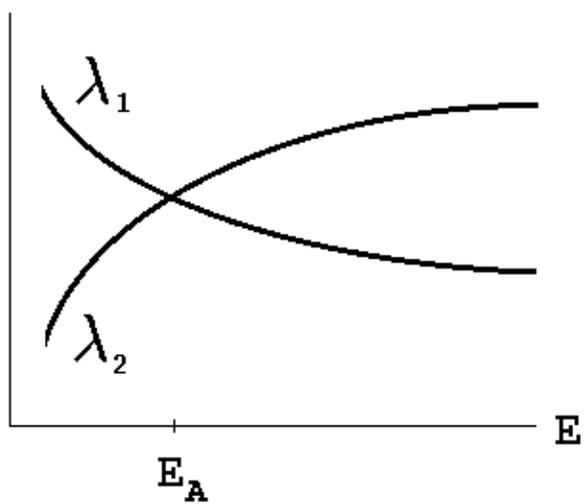
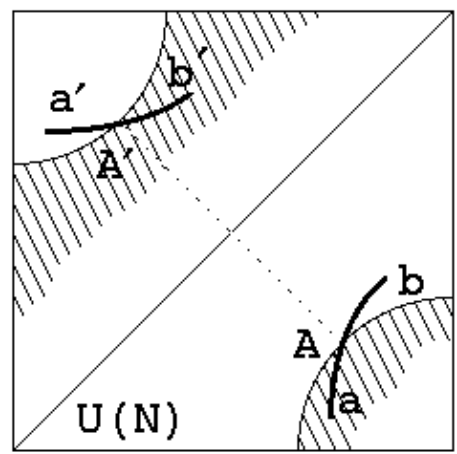
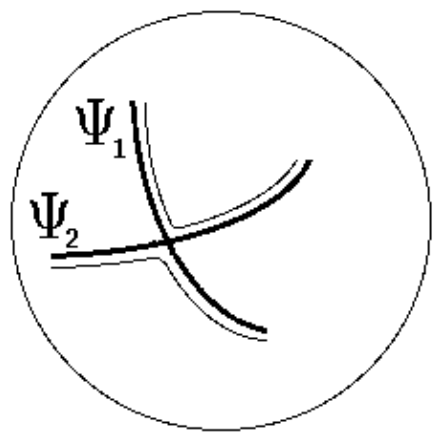
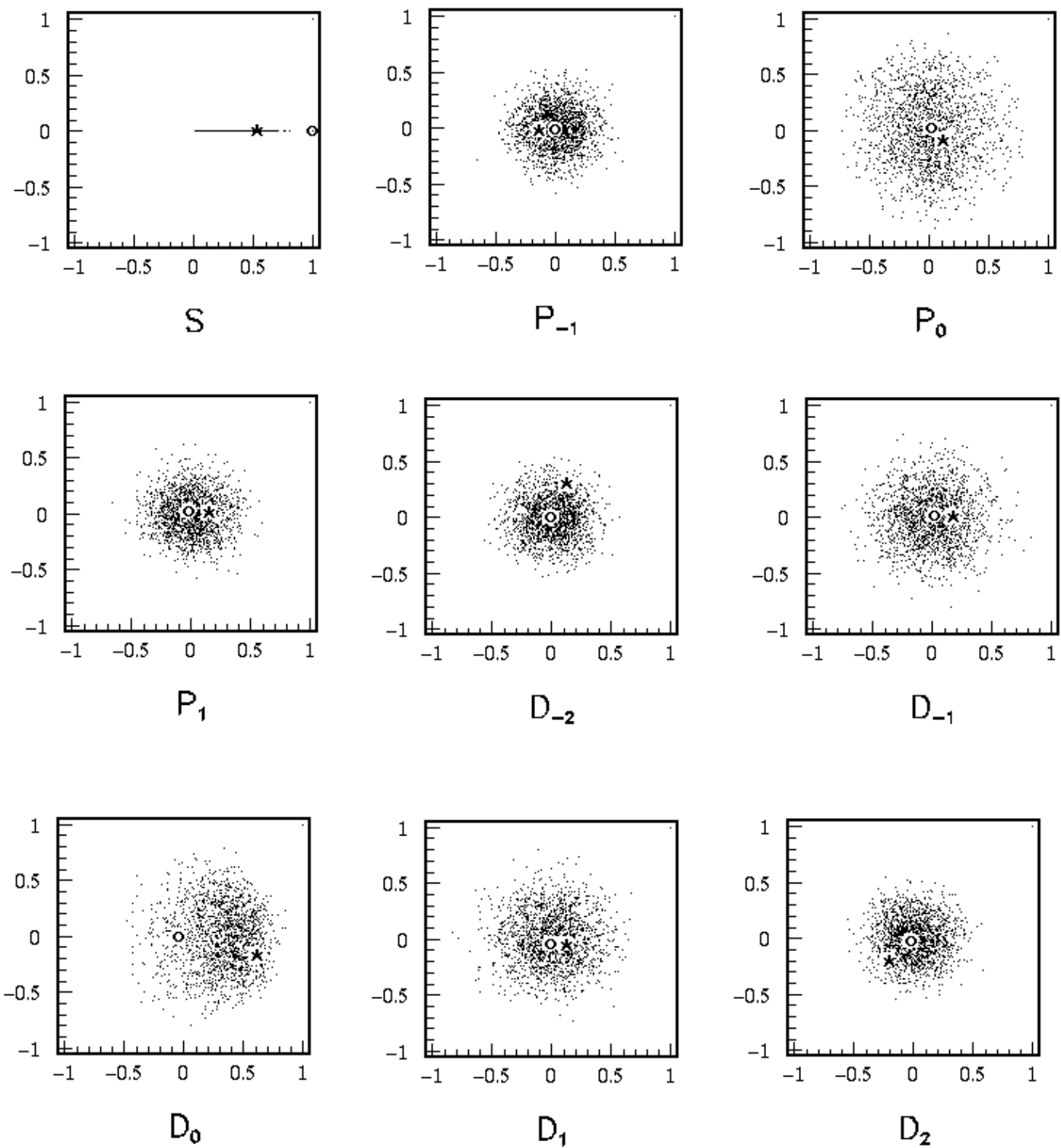
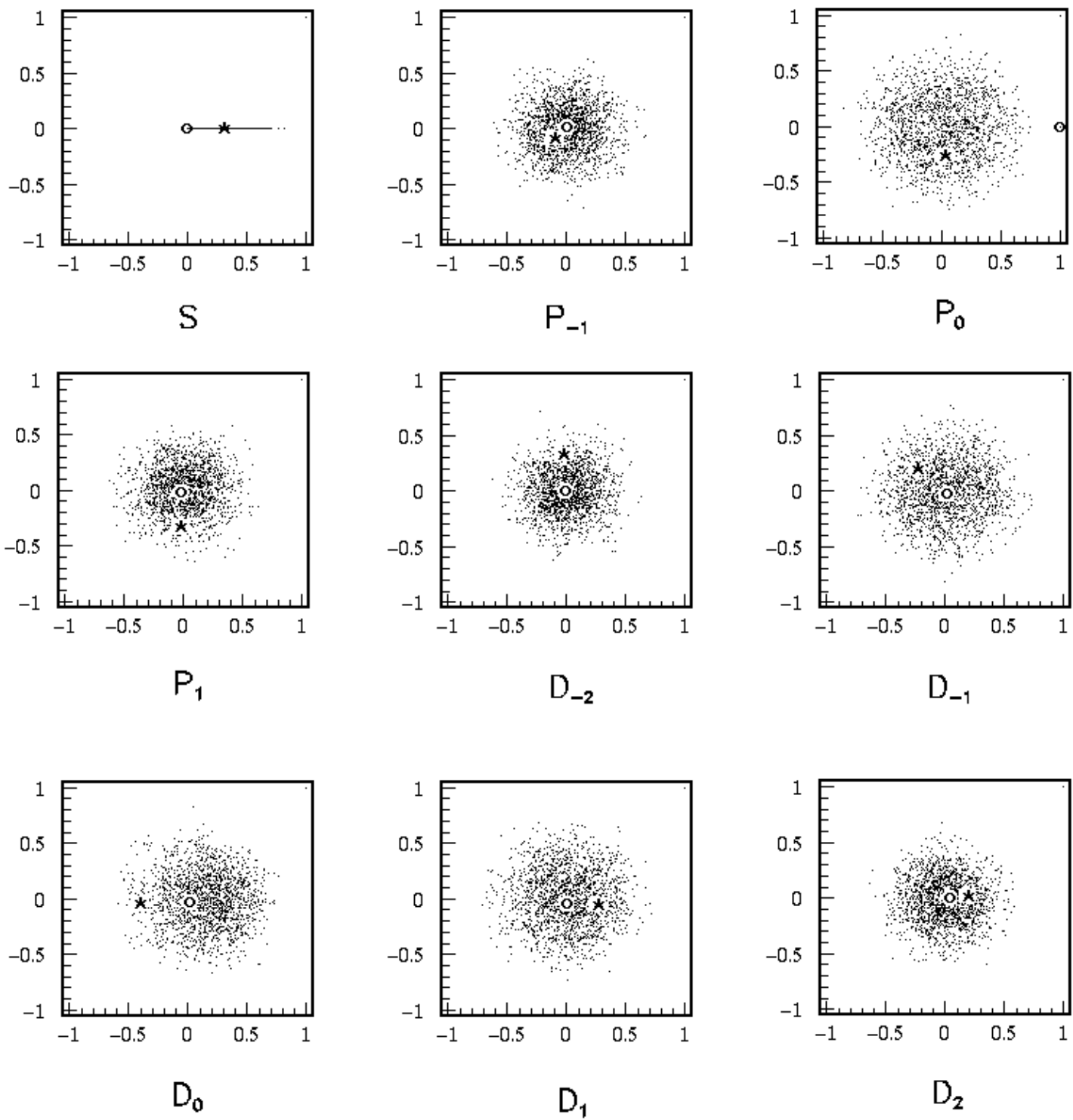
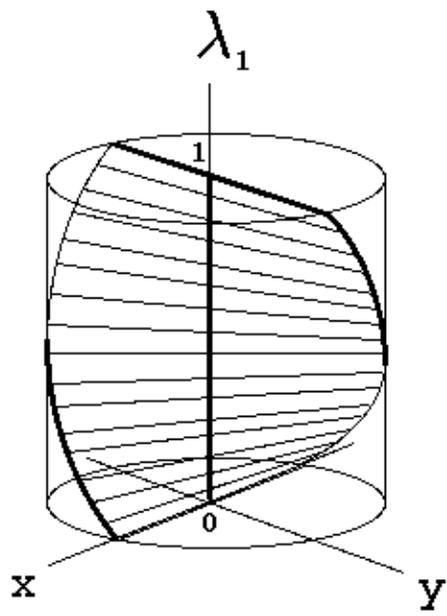
\end{document}